\documentclass[10pt,onecolumn,draftcls,a4paper]{IEEEtran}
\usepackage{graphicx}

\usepackage{amsmath}
\usepackage{amssymb}

\usepackage{tikz}
\usetikzlibrary{arrows}
\usepackage{pgf}
\usepackage{schemabloc} 

\usepackage{multirow}

\usepackage{multicol}
\usepackage{fixltx2e}
\usepackage{subfig}

\usepackage[margin=0.8in]{geometry}

\begin{document}
\title{On the Convergence Speed of Turbo Demodulation with Turbo Decoding}

\author{Salim~Haddad,
        Amer~Baghdadi,
        and~Michel~Jezequel

%\thanks{The authors are with the Electronics Department
%of Telecom Bretagne, Institut Telecom, Lab-STICC CNRS, Technopole Brest-Iroise CS 
%83818, 29238 Brest Cedex 3. e-mail: \{salim.haddad,amer.baghdadi,michel.jezequel\}@telecom-bretagne.eu.}% <-this % stops a space
}

\markboth{IEEE Transactions on Signal Processing}
%\markboth{Journal of \LaTeX\ Class Files,~Vol.~6, No.~1, January~2007}%
{Shell \MakeLowercase{\textit{et al.}}: Bare Demo of IEEEtran.cls for Journals}

\maketitle

%\vspace{-20pt}
\begin{abstract}
%\boldmath
Iterative processing is widely adopted nowadays in modern wireless receivers for advanced channel codes like turbo and LDPC codes. Extension of this principle with an additional iterative feedback loop to the demapping function has proven to provide substantial error performance gain. However, the adoption of iterative demodulation with turbo decoding is constrained by the additional implied implementation complexity, heavily impacting latency and power consumption. In this paper, we analyze the convergence speed of these combined two iterative processes in order to determine the exact required number of iterations at each level. Extrinsic information transfer (EXIT) charts are used for a thorough analysis at different modulation orders and code rates. An original iteration scheduling is proposed reducing two demapping iterations with reasonable performance loss of less than 0.15 dB. Analyzing and normalizing the computational and memory access complexity, which directly impact latency and power consumption, demonstrates the considerable gains of the proposed scheduling and the promising contributions of the proposed analysis.  
\end{abstract}

\vspace{-2pt}
%\begin{IEEEkeywords}
%Iterative Demodulation, Turbo Decoding, Convergence speed, Complexity, Arithmetic Operations, Memory access, EXIT charts
%\end{IEEEkeywords}

\IEEEpeerreviewmaketitle

\section{Introduction}
\IEEEPARstart{A}{dvanced} wireless communication standards impose the use of modern techniques to improve spectral efficiency and reliability. Among these techniques Bit-Interleaved Coded Modulation (BICM) with different modulation orders and Turbo Codes with various code rates are frequently adopted. 

The BICM principle \cite{caire_bit-interleaved_1997} currently represents the state-of-the-art in coded modulations over fading channels. The Bit-Interleaved Coded Modulation with Iterative Demapping (BICM-ID) scheme proposed in \cite{xiaodong_li_bit-interleaved_1997} is based on BICM with additional soft feedback from the Soft-Input Soft-Output (SISO) convolutional decoder to the constellation demapper. In this context, several techniques and configurations have been explored. In \cite{chindapol_design_2001}, the authors investigated different mapping techniques suited for BICM-ID and QAM16 constellations. They proposed several mapping schemes providing significant coding gains. In \cite{Abramovici_Onturbo_1999}, the convolutional code classically used in BICM-ID schemes was replaced by a turbo code. Only a small gain of 0.1 dB was observed.  This result may make BICM-ID with turbo-like coding solutions (TBICM-ID) unsatisfactory with respect to the added decoding complexity. On the other hand, authors in \cite{nour_cth11-4:loweringerror_2006} have presented a technique intended to improve the performance of TBICM-ID over non Gaussian channels. The proposed technique, namely Signal Space Diversity (SSD), consists of a rotation of the constellation followed by a signal space component interleaving. It has shown additional error correction at the receiver side in an iterative processing scenario. 

Constellation rotation enables to exploit higher code rates and to solve potential problems in selective channels while keeping good performance. It has been proposed for all constellation orders of Quadrature Amplitude Modulation (QAM). Combining constellation rotation with signal space component interleaving leads to significant improvement in performance over fading and erasure channels. It increases the diversity order of a communication system without using extra bandwidth. BICM coupled with SSD has been extensively studied for single carrier systems, e.g. in \cite{kiyani_iterative_2007,boutros_signal_1998} and the references therein.

Application of iterative demapping in this context has shown excellent error rate performance results particularly in severe channel conditions (erasure, multi-path, real fading models) \cite{abdel_nour_improving_2008}. In that work, LDPC channel coding was considered.

Nevertheless, most of the existing works have not considered these techniques from an implementation perspective. In fact, the application of the iterative demapping in future receivers using advanced iterative channel decoding will lead to further latency problems, more power consumption and more complexity caused by feedback inner and outer the decoder. Besides extrinsic information exchange inside the iterative channel decoder, additional extrinsic information is fed back as {\it a priori} information used by the demapper to improve the symbol to bit conversion. The number of iterations to be run at each level should be determined accurately as it impacts significantly, besides error rate performance, latency, power consumption, and complexity. 

This work discusses the implementation efficiency of iterative receiver based on turbo demodulation and turbo decoding in order to achieve a gain in band-limited wireless communication systems. Convergence speed is analyzed for various system configurations to determine the exact required number of iterations at each level. Significant complexity reductions can be achieved by means of the proposed original iteration scheduling. Finally, an accurate normalized complexity analysis is presented in terms of arithmetic and memory access operations.

%The rest of the paper is organized as follows. Section \ref{SystemModelSection} presents the system model with the associated parameters and gives a brief description of the underlined algorithms for turbo demapping and turbo decoding. Section \ref{ConvSpeedAnalysis} illustrates the impact of constellation rotation and iterative demapping on the convergence speed using EXIT chart based analysis. Section \ref{RedDemIterSection} presents the proposed iteration scheduling and evaluates its impact on the error rate performance for different modulation orders and code rates. Section \ref{ComplexityEvalNormaliz} presents an evaluation of the receiver complexity  in terms of number and type of arithmetic operations and memory access. Section \ref{DiscussionSection} discusses the achieved complexity reductions. Finally, section \ref{Conclusion} concludes the paper.

 
%\hfill April 11, 2011
\vspace{-2pt}
\section{System model and Algorithms}
\label{SystemModelSection}
%This section details the overall system model and considered parameters of the transmitter, channel, and receiver. In addition, it gives a brief presentation of the underlined algorithms for the iterative demapping and decoding.

\vspace{-5pt}
\subsection{System Model}
\label{Section2A}
\begin{figure}[h!]
\begin{center}
\scalebox{0.42}{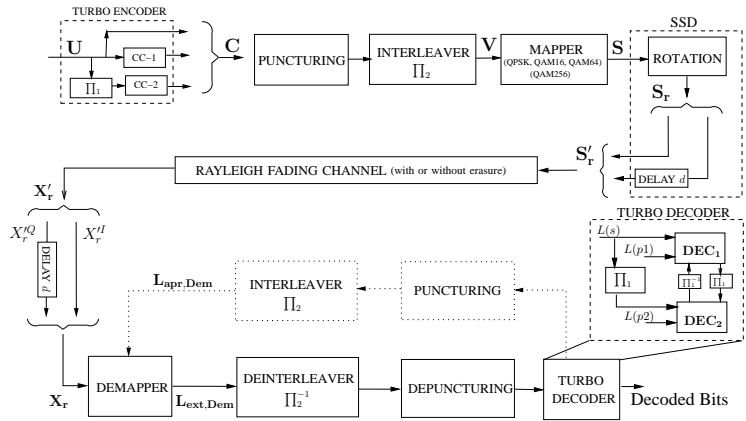}
\caption{\label{SysMod} System model with TBICM-ID-SSD.}
\end{center} \vspace{-20pt}
\end{figure}

The considered system uses one transmit and one receive antenna while assuming perfect synchronization. Fig. \ref{SysMod} shows a basic transmitter and receiver model using turbo demodulation and decoding. We denote by TBICM-ID-SSD a turbo BICM with iterative demapping coupled with signal space diversity.

On the transmitter side, information bits $U$ which are called systematic bits are regrouped into symbols $u_p$ consisting of $l$ bits, and encoded with an $l$-binary turbo encoder. It consists of a parallel concatenation of two identical convolutional codes (PCCC). The output codeword $C$ is then punctured to reach a desired coding rate $R_c$. The $8$-state double binary ($l=2$) recursive systematic convolutional code (RSC code), adopted in the WiMax standard, is considered for the turbo encoder.

%%%%%%%%%%%%%%%Fig. \ref{TurboEncoderDouble} shows the considered $8$-state double binary ($l=2$) recursive systematic convolutional code (RSC code), adopted in the WiMax standard.

%\begin{figure}[h!]
%\begin{center}
%\scalebox{0.48}{\input{encoders.pstex_t}}
%\caption{\label{TurboEncoderDouble} 8-state double binary RSC.}
%\end{center} 
%\end{figure}

In order to gain resilience against error bursts, the resulting sequence is interleaved using an $S$-random interleaver $\Pi_2$ with $S=\sqrt{N/4}$. Punctured and interleaved bits denoted by $V$ are then gray mapped to channel symbols $s_q$ chosen from a  $2^M$-ary constellation $X$, $M$ is the number of bits per modulated symbol.

%In order to gain resilience against error bursts \cite{koutsouvelis_low_2008}, the resulting sequence is interleaved using an $S$-random interleaver $\Pi_2$ with $S=\sqrt{N/4}$. The $S$-random interleaver guarantees that no input symbols within distance $S$ appear within a distance of $S$ in the output. Punctured and interleaved bits denoted by $V$ are then gray mapped to complex channel symbols $s_q$ chosen from a  $2^M$-ary constellation $X$, where $M$ is the number of bits per modulated symbol.

%%%%%%%%%%%%%%%%%Gray mapping produces the best results \cite{jun_tan_analysis_2005} for TBICM. Restricted to Gray mapping, QAM schemes offer different types of bit protection depending on the position of the allocated bit within the transmitted symbol and the order of the modulation. For example, the Gray mapped QAM16 of Fig. \ref{QAM16ConstRot} offers two levels of bit protection. $c_0$ and $c_1$ can provide more protection than $c_2$ and $c_3$. Bit error rates at the latter two bits is approximately double that of the first two bits. In \cite{le_goff_turbo-codes_1994}, the most protected bit positions were allocated to systematic bits. In \cite{Barbulescu97bandwidthefficient}, these positions were allocated to parity bits. The effect of bit protection positions on the convergence speed will be analyzed in the next section.

%\begin{figure}[h!]
%\begin{center}
%\scalebox{0.3}{\input{QAM16ConstRot.pstex_t}}
%\caption{QAM16 gray mapped constellation with rotation.}
%\label{QAM16ConstRot}
%\end{center}  
%\end{figure}

Applying the SSD consists first of the rotation of the mapped symbols $s_q$. The resulting rotated symbols are denoted by $s_{r,q}$. The performance gain obtained when using a rotated constellation $X_r$ depends on the choice of the rotation angle. In this regard, a thorough analysis has been done for the 2nd-generation terrestrial transmission system developed by the DVB Project (DVB-T2) which adopted the rotated constellation technique. A single rotation angle \cite{abdel_nour_improving_2008} has been chosen for each constellation size independently of the channel type. Using these angles, and with LDPC code, gains of 0.5 dB and 6 dB were shown for Rayleigh fading channel without and with erasure respectively for high code rate \cite{abdel_nour_improving_2008}. The second step consists of signal space component interleaving. When a constellation signal is submitted to a fading event, its in-phase component $I$ and quadrature component $Q$ fade identically and suffers from an irreversible loss. A means of avoiding this loss involves making $I$ and $Q$ fade independently while each carrying all the information regarding the transmitted symbol. By inserting an interleaver (a simple delay for uncorrelated fast-fading channel) between the $I$ and $Q$ channels, the diversity order is doubled.

$s'_{r,q}$ symbols are then transmitted over a noisy and Rayleigh fast-fading channel with or without erasures. \textcolor{black}{In fact, erasure events happen in single frequency network (SFN), as in DVB-T2 standard, due to destructive interferences.} Each received symbol $x'_{r,q}$ is affected by a different fading coefficient, an erasure coefficient, and additive Gaussian noise.

The channel model considered is a frequency non-selective memoryless channel with erasure probability. The received discrete time baseband complex signal can be written as:

\vspace{-13pt}
\begin{eqnarray}
\label{h_q}
x_{r,q}' & = & h_q . \rho_q . s_{r,q}' + n_q  \nonumber \\ 
 & = & h'_q . s_{r,q}' + n_q
\end{eqnarray}

where $h_q$ is the Rayleigh fast-fading coefficient, $\rho_q$ is the erasure coefficient probability taking value 0 with a probability $P_\rho$ and value 1 with a probability of $1-P_\rho$. $n_q$ is a complex white Gaussian noise with spectral density $N_0/2$ in each component axes, and $h'_q$ is the channel attenuation.  Note that, at the receiver side, the transmitted energy has to be normalized by a $\sqrt{1-P_\rho}$ factor in order to cope with the loss of transmitted power. 

\vspace{-5pt}
\subsection{Max-Log-MAP Demapping Algorithm}
\label{DemapAlg}
%At the receiver side, complex received symbols $x_{r,q}'$ are Q-components re-shifted and demapped. 
At the receiver side, the complex received symbols $x_{r,q}'$ have their Q-components re-shifted resulting in $x_{r,q}$. An extrinsic log-likelihood ratio $L_{ext,Dem}(c_{p,q}/x_{r,q})$ is calculated for each bit $c_{p,q}$ corresponding to the $p^{th}$ bit of the received rotated and modulated symbol $x_{r,q}$. After de-interleaving, de-puncturing and turbo decoding, extrinsic information from the turbo decoder $L_{ext,Dec}(c_{p,q})$ is passed through the interleaver, punctured and fed back as {\it a priori} information $L_{apr,Dem}(c_{p,q})$ to the demapper in a turbo demapping scheme. $L_{ext,Dem}(c_{p,q}/x_{r,q})$ is the difference between the soft output {\it a posteriori} $L_{Dem}(c_{p,q}/x_{r,q})$ and $L_{apr,Dem}(c_{p,q})$ at the demapper side, it is given by the expression below: \vspace{-10pt}

%\vspace{-3pt}
\begin{eqnarray} 
 L_{ext,Dem}(c_{p,q}/x_{r,q}) &=& L_{Dem}(c_{p,q}/x_{r,q})-L_{apr,Dem}(c_{p,q}) \nonumber \\
  &=& \log \left( \frac{Z_1}{Z_2} \right)
\label{DemapperEquationMAP}
\end{eqnarray}
$Z_{l(l=0,1)}$ can be expressed as:
\begin{eqnarray}
%Z_{l(l=0,1)}=\sum\limits_{s_{r,j} \in X^k_{r,l}} e^{-\frac{|x_{r,q}-h_q s_{r,j}|^2}{\sigma^2}}. \prod\limits_{i=0,i \not=k }^{M-1} P(c_{i,q})
Z_{l(l=0,1)}=\sum\limits_{s_{r,j} \in X^k_{r,l}} e^{-A_q} . \prod\limits_{i=0,i \not=k }^{M-1} P(c_{i,q})
\end{eqnarray}
where $X^k_{r,l}$, with $l \in \{0,1\}$, are the symbol sets of the constellation for which symbols have their $i^{th}$ bit equal to $l$. $P(c_{i,q})$ is the probability of the $i^{th}$ bit of constellation symbol $s_{r,q}$ computed through {\it a priori} information $L_{apr,Dem}(c_{i,q})$. Reducing the complexity of the expressions above can be performed by applying the max-log approximation. Thus, equation (\ref{DemapperEquationMAP}) can be written as: \vspace{-5pt}

%{\tiny \begin{align}
%L_{ext,Dem}(c_{k,q}/x_{r,q}) = \min\limits_{S_{r,j} \in X^k_{r,0}} \left(\frac{|x^I_{r,q-1}-h_{q-1} S^I_{r,j}|^2+|x^Q_{r,q-1}-h_{q-1} S^Q_{r,j}|^2}{2 \sigma^2}-\sum\limits^{M-1}_{i=0,i\not=k,c_{i,q}=1} L_{apr,Dem}(c_{i,q})\right) \nonumber \\ - \min\limits_{S_{r,j} \in X^k_{r,1}} \left(\frac{|x^I_{r,q-1}-h_{q-1} S^I_{r,j}|^2+|x^Q_{r,q-1}-h_{q-1} S^Q_{r,j}|^2}{2 \sigma^2}-\sum\limits^{M-1}_{i=0,i\not=k,c_{i,q}=1} L_{apr,Dem}(c_{i,q})\right)
%\label{DemapperEquationMaxLogMAP}
%\end{align}}

\begin{equation}
L_{ext,Dem}(c_{p,q}/x_{r,q}) = \min\limits_{s_{r,j} \in X^k_{r,0}} (A_q-B_{p,q}) - \min\limits_{s_{r,j} \in X^k_{r,1}} (A_q-B_{p,q})
\label{DemapperEquationMaxLogMAP}
\end{equation}
where $A_q$ and $B_{p,q}$ are computed as follows.
\begin{eqnarray}
A_q=\frac{h'^2_{q}}{\sigma^2} |x^I_{r,q}- s^I_{r,j}|^2 + \frac{h'^2_{q-1}}{\sigma^2} |x^Q_{r,q}-s^Q_{r,j}|^2 \\
\label{APrioriAdder}
B_{p,q}=\left (\sum\limits^{M-1}_{i=0,c_{i,q}=1} L_{apr,Dem}(c_{i,q}) \right)-L_{apr,Dem}(c_{p,q})
\end{eqnarray}
\textcolor{black}{In fact, the above demapping equations are valid for both channel models (with or without erasures) through the use of $h'_q$ coefficient (equation (\ref{h_q})).}

These simplified expressions exhibit three main computation steps: (a) Euclidean distance computation referred by $A_q$, (b) {\it a priori} adder referred by $B_{p,q}$, and (c) minimum finder referred by the $min$ operation of equation (\ref{DemapperEquationMaxLogMAP}). 

\vspace{-5pt}
\subsection{Max-Log-MAP Decoding Algorithm}
Following the demapping function at the receiver side, the turbo decoding algorithm is applied. The BCJR algorithm is considered for the Soft Input Soft Output (SISO) convolutional decoders. Using input symbols and {\it a priori} extrinsic information, each SISO decoder computes { \it a posteriori} probabilities. The BCJR SISO decoder computes first the branch metrics $\gamma$. Then it computes the forward $\alpha_{k}$ and backward $\beta_{k}$ metrics between two trellis states $s$ and $s'$ \cite{muller_exploring_2006}. 
\begin{equation}
  \label{AlfaComput}
	\alpha_{k}(s) = \displaystyle\max_{(s^{'},s)} (\alpha_{k-1}(s^{'}) + \gamma_{k}(s^{'},s))
\end{equation}
\begin{equation}
  \label{BetaComput}
	\beta_{k}(s) = \displaystyle\max_{(s^{'},s)} (\beta_{k+1}(s^{'}) + \gamma_{k+1}(s^{'},s))
\end{equation}
where 
\begin{equation}
  \label{eq:5.3}
	\gamma_{k}(s^{'},s) = \gamma^{Sys}_{k}(s^{'},s)+\gamma^{Parity}_{k}(s^{'},s)+\gamma^{Ext}_{k}(s^{'},s)
\end{equation}

The soft output information $so(d_{k}=c_pc_{p+1})$ and symbol-level extrinsic information $z(d_{k}=c_pc_{p+1})$ of symbol $k$ are then computed using equations (\ref{SoftOutput}) and (\ref{ExtrinsicSymbOutput}). The extrinsic information, which is exchanged iteratively between the two SISO decoders, is obtained by subtracting the intrinsic information from $so(d_{k}=c_pc_{p+1})$.

\vspace{-13pt}
\begin{equation}
  \label{SoftOutput}
	so(d_{k})=\displaystyle\max_{(s^{'},s)/d(s^{'},s)=d_k} \left(\alpha_{k-1}(s^{'})+\gamma_{k}(s^{'},s)+ \beta_{k}(s)\right) 
\end{equation}
\begin{equation}
  \label{ExtrinsicSymbOutput}
	z(d_{k})=\displaystyle\max_{(s^{'},s)/d(s^{'},s)=d_k} \left(\alpha_{k-1}(s^{'})+\gamma^{Ext}_{k}(s^{'},s)+ \beta_{k}(s)\right) 
\end{equation}

$z(d_{k})$ can be multiplied by a constant scaling factor $SF$ (typically equals to 0.75) for a modified Max-Log-MAP algorithm improving the resultant error rate performance.

Finally, in case of turbo demapping and only by one SISO decoder, the bit-level extrinsic information of systematic symbols $c_{p}c_{p+1}$ are computed using equations (\ref{ExtrinsicBitOutput1}) and (\ref{ExtrinsicBitOutput2}). Similar computations are done for parity symbols $c_{p+2}c_{p+3}$.

\vspace{-5pt}
\begin{equation}
\label{ExtrinsicBitOutput1}
L_{apr,Dem}(c_p) = \displaystyle\max_{} [z(d_{k}=11),z(d_{k}=10)] - \displaystyle\max_{}[z(d_{k}=01),z(d_{k}=00)]
\end{equation}
\begin{equation}
\label{ExtrinsicBitOutput2}
L_{apr,Dem}(c_{p+1}) = \displaystyle\max_{}[z(d_{k}=11),z(d_{k}=01)] - \displaystyle\max_{}[z(d_{k}=10),z(d_{k}=00)]
\end{equation}

%where $SF$ is the constant scale factor for the decoding Max-Log-MAP algorithm.

These expressions exhibit three main computation steps: (a) branch metrics computation referred by $\gamma_k$, (b) state metrics computation referred by ($\alpha_k$ and $\beta_k$), and (c) extrinsic information computation referred by $L_{apr,Dem}$ and $z$.

\section{TBICM-SSD and TBICM-ID-SSD convergence speed analysis}
\label{ConvSpeedAnalysis}
\textcolor{black}{This section illustrates the impact of constellation rotation and iterative demapping on the convergence speed. Convergence speed designates the rapidity of the convergence of the iterative process.}

\textcolor{black}{EXIT charts \cite{ten_brink_convergence_1999} are used as a useful tool for a clear and thorough analysis of the convergence speed. They were first proposed for parallel concatenated codes, and then extended to other iterative processes. For iterative demapping with turbo decoding (TBICM-ID-SSD), authors in \cite{nour_cth11-4:loweringerror_2006} have used this tool to analyze the iterative exchange of information between the different SISO components. In this system receiver with two iterative processes, the response of the two SISO decoders is plotted while taking into consideration the SISO demapper with updated inputs and outputs.}

\textcolor{black}{In this scheme, $IA_1$, $IA_2$, $IE_1$, $IE_2$ are used to designate the {\it a priori} and extrinsic information respectively for $\mathtt{DEC_1}$ and $\mathtt{DEC_2}$ (Fig. \ref{SysMod}). Iterations  start  without {\it a priori} information ($IA_1 =0$  and  $IA_2 =0$). Then, extrinsic information $IE_1$ of $\mathtt{DEC_1}$ is fed to $\mathtt{DEC_2}$ as {\it a priori} information $IA_2$ and vice versa, i.e. $IE_1=IA_2$ and $IE_2=IA_1$. Since this EXIT chart analysis is asymptotic, infinite long BICM interleaver size should be assumed. The SISO decoder is represented by its transfer function:}

\vspace{-15pt}
\begin{equation}
IE=T(IA,E_b/N_0) \vspace{-5pt}
\end{equation}

\begin{figure}[h!]
\begin{center}
\scalebox{0.36}{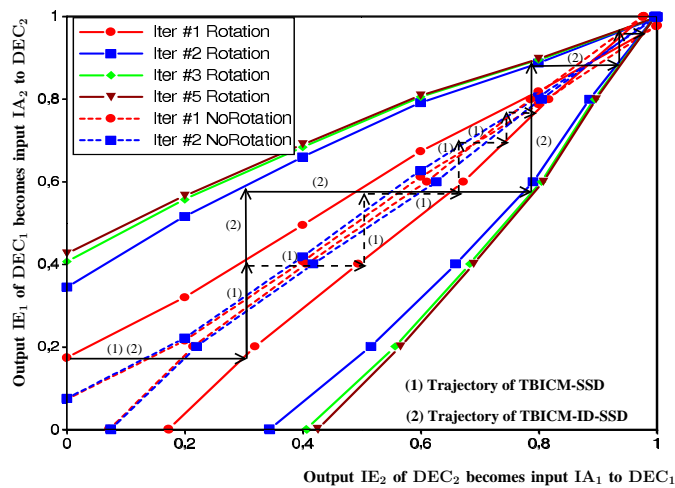}
\caption{\label{EffectRotationWithErasureEXIT} EXIT chart analysis at an $E_b/N_0$= 22 dB of the dual binary turbo decoder for iterations to the QAM64 demapper. Code rate $\frac{4}{5}$ is considered for transmission over Rayleigh fast-fading channel with erasure probability equals to 0.15.}
\end{center} \vspace{-15pt}
\end{figure}

\textcolor{black}{Extensive analysis for different $E_b/N_0$ and different system parameters (modulation orders and code rates) has been conducted and gave similar results. Fig. \ref{EffectRotationWithErasureEXIT} illustrates one of these simulations for QAM64, code rate $\frac{4}{5}$, $E_b/N_0$=22 dB, and erasure probability equals to 0.15. The transfer function of the turbo decoder is represented by the two-dimensional chart as follows. One SISO decoder component is plotted with its input on the horizontal axis and its output on the vertical axis. The other SISO component is plotted with its input on the vertical axis and its output on the horizontal axis. The iterative decoding corresponds to the trajectory found by stepping between the two curves. For a successful decoding, there must be a clear path between the two curves so that iterative decoding can proceed from 0 to 1 mutual extrinsic information.} 

\textcolor{black}{The plain curves of Fig. \ref{EffectRotationWithErasureEXIT} correspond to the EXIT charts for the case with rotated constellation. Meanwhile, the dashed curves correspond to the case with no rotation. Furthermore, the red curves correspond to non iterative demapping, e.g. SISO demapper executed once in feed forward scheme. Applying demapping iterations corresponds to the other colored curves in the EXIT charts of Fig. \ref{EffectRotationWithErasureEXIT}.}

\textcolor{black}{In this figure, we observe that the EXIT tunnel is wider for the rotated case than the one without. Furthermore, the tunnel is limited to that of one demapping iteration for the latter case. Thus, making more demapping iterations will not affect the convergence speed of non rotated constellation configurations. However the tunnel is enlarging (improving) until three demapping iterations using the rotated constellation. For TBICM-SSD, EXIT charts show a need of more than 6 turbo decoding iterations to attain convergence following the trajectory (1). Whereas 4 demapping iterations are sufficient following the trajectory (2).}

\textcolor{black}{Thus, in case of TBICM-ID-SSD, the iteration scheduling which optimize the convergence is the one that enlarge the EXIT tunnel as soon as possible. Analyzing the different tunnel curves in the EXIT figure shows that the tunnel is enlarging for each demapping iteration. Thus, the optimized scheduling is to execute only one turbo decoding iteration for each demapping iteration and then step forward to the next demapping iteration (enlarge the EXIT tunnel). This scheduling is the one adopted implicitly in \cite{nour_cth11-4:loweringerror_2006}. Note that after the third demapping iteration, only a slight improvement in convergence is observed. Similar results have been found for all considered modulation orders, code rates and erasure coefficients. 
This result will be used in the next section to reduce the number of demapping iterations.}

\section{Reducing the number of demapping iterations}
\label{RedDemIterSection}
As mentioned in the previous section, the optimized profile applies one turbo code iteration for each demapping iteration. Thus, reducing the number of turbo demapping iterations will reduce the total number of iterations for the turbo decoder. 

However, various constructed EXIT charts with different parameters show that after a specific number of demapping iterations, only a slight improvement is predicted. As an example, in Fig. \ref{EffectRotationWithErasureEXIT} decoder transfer functions coincide with each other after 3 demapping iterations. However, one can notice that turbo decoding iterations must continue until that the two constituent decoders agree with each other. Thus, the number of demapping iterations can be reduced without affecting error rates, while keeping the same total number of turbo decoding iterations. This constitutes the basis for our proposed original iteration scheduling.

\begin{figure}[h!]
\center
\includegraphics[clip, scale=0.55]{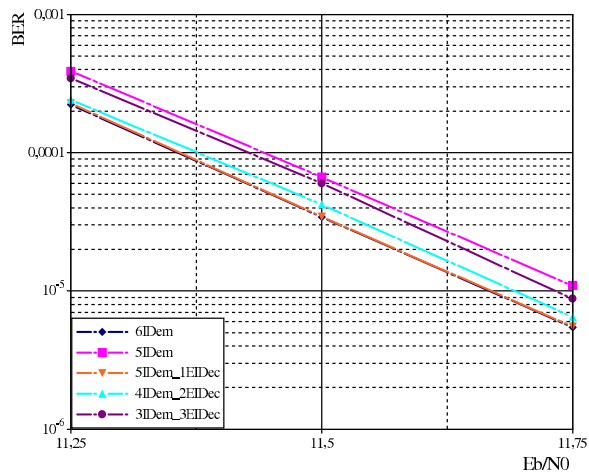}
\caption{BER performance comparison for TBICM-ID-SSD for the transmission of 1536 information bits frame over Rayleigh fast-fading channel without erasure. QAM64 modulation scheme with code rate $\frac{2}{3}$ are considered.}
\label{OptimTBICMIDSSDQAM64_0.67} \vspace{-5pt}
\end{figure}

In fact, to keep the same number of iterations for the decoder unaltered, one turbo code iteration is added after the last iteration to the demapper for each eliminated demapping iteration. Fig. \ref{OptimTBICMIDSSDQAM64_0.67} simulates six turbo demapping iterations performing one turbo decoding iteration for each. Hence, six turbo code iterations are performed in total. This scheme is denoted as $6IDem$. 

With the proposed iteration scheduling, $5IDem\_1EIDec$ designates five demapping iterations (one turbo code iteration is applied for each) followed by one extra turbo code iteration. 

Referring to Fig. \ref{OptimTBICMIDSSDQAM64_0.67}, error rates associated to $6IDem$ and $5IDem\_1EIDec$ show almost same performances, while one feedback to the demapper is eliminated in the latter scheme. Similarly, for $4IDem\_2EIDec$, two feedbacks to the demapper are eliminated. A slight loss of 0.025 dB is induced. Eliminating more demapping iterations will cause significant performance degradation. $3IDem\_3EIDec$ is closer to $5IDem$ than to $6IDem$. 

%\begin{figure}[h]% 
%\centering
%\subfloat[]{%
%\label{OptimTBICMIDSSDQAM64_0.67}
%\includegraphics[width=8cm]{Figures/OptimTBICMIDSSDQAM64_0.67.eps}}%  
%\hspace{0.5pt}%
%\subfloat[]{%
%\label{OptimTBICMIDSSDQAM16_0.8_0.15}%
%\includegraphics[width=8cm]{Figures/OptimTBICMIDSSDQAM16_0.8_0.15.eps}}\\  
%\caption[A set of four sub-floats.]{BER performance comparison for TBICM-ID-SSD for the transmission of 1536 information bits frame over Rayleigh fading channel
%\subref{OptimTBICMIDSSDQAM64_0.67} Without erasure, QAM64 modulation scheme with code rate $\frac{2}{3}$ is considered.
%\subref{OptimTBICMIDSSDQAM16_0.8_0.15} With erasure probability equals to 0.15, QAM16 modulation scheme with code rate $\frac{4}{5}$ is considered.}    
%\label{OptimTBICMIDSSD}%
%\end{figure}

\begin{table}[h!] \vspace{-2pt}
\centering
\scalebox{0.9}{
\begin{tabular}{|c|c|c|}
\hline
Modulation scheme & \multicolumn{2}{c|}{Performance loss (dB)} \\
\cline{2-3}

 & Without Erasure & With Erasure \\
 & $R_c=6/7$ $-\triangleright$ $R_c=1/2$ & $R_c=6/7$ $-\triangleright$ $R_c=1/2$ \\

\hline
QPSK & 0.02 $-\triangleright$ 0.03 & 0.02 $-\triangleright$ 0.05 \\
\hline
QAM16 & 0.04 $-\triangleright$ 0.06 & 0.04 $-\triangleright$ 0.08 \\
\hline
QAM64 & 0.05 $-\triangleright$ 0.08 & 0.07 $-\triangleright$ 0.12 \\
\hline
QAM256 & 0.07 $-\triangleright$ 0.10 & 0.09 $-\triangleright$ 0.15 \\
\hline
\end{tabular}
}
\caption{Performance loss for different modulation schemes and code rates after 2 omitted demapping iterations.}
\label{BERLostReduction} \vspace{-20pt}
\end{table}

In fact, $4IDem\_2EIDec$ represents the most optimized curve for the $6IDem$ performance scheme as shown in Fig. \ref{OptimTBICMIDSSDQAM64_0.67}. EXIT charts do not agree with this consideration at the first sight, three demapping iterations were sufficient to do the same correction as the eight iterations. EXIT charts are based on average calculations as many frames are simulated. The three demapping iterations represents the average number of demapping iterations needed to be sure that the two constituent decoders agree with each other. Making more demapping iterations will provide more error correction. Further simulations show performance loss of 0.02 dB to 0.1 dB and 0.02 dB to 0.15 dB for no erasure and erasure events respectively when the proposed scheduling is applied. Table \ref{BERLostReduction} summarizes the reduced performance loss for different code rates and constellation orders after omitting two demapping iterations. \textcolor{black}{These values were investigated for the worst case corresponding to $3IDem\_2EIDec$ in comparison to $5IDem$.} \textcolor{black}{Note that for error floor region, simulations show almost identical BER performance if applying more than 3 demapping iterations.} \textcolor{black}{Furthermore, it is worth noting that with a limited-diversity channel model, omitting 2 demapping iterations leads to slightly lower performance loss than those of Table \ref{BERLostReduction} for a fast-fading channel model. In fact, one demapping iteration with high-diversity channel model leads to more error correction compared to one iteration executed with limited-diversity one. Conducted simulations with block-fading channel model have confirmed this result.} 

Using this technique, latency and complexity issues caused by the TBICM-ID-SSD are reduced. Two feedbacks to the demapper with the associated delays, computations, and memory accesses are eliminated. \textcolor{black}{It is worth noting that the proposed new scheduling does not have any impact on the receiver area (logic or memory). This scheduling is applied on a TBICM-ID-SSD receiver and proposes complexity reduction in ``temporal dimension'' (which impacts power consumption, throughput, and latency).} Complexity reductions will be evaluated and discussed in the next section.

\section{Complexity evaluation and normalization}
\label{ComplexityEvalNormaliz}
The main motivation behind the conducted convergence speed analysis and the proposed technique for reducing the number of iterations is to improve the receiver implementation quality. In order to appreciate the achieved improvements, an accurate evaluation of the complexity in terms of number and type of operations and memory access is required. \textcolor{black}{Such complexity evaluation is fair and generalized as it is independent from the architecture mode (serial or parallel) and remains valid for both of them. In fact, all architecture alternatives should execute the same number of operations (serially or concurrently) to process a received frame.} In this section, we consider the two main blocks of the TBICM-ID-SSD system configuration which are the SISO demapper and the SISO decoder. The proposed evaluation considers the low complexity algorithms presented in section \ref{SystemModelSection}. A typical fixed-point representation of channel inputs and various metrics is considered. Table \ref{OperBit} summarizes the total number of required quantization bits for each parameter.

\begin{table}
\parbox{.5\linewidth}{

\scalebox{0.9}{
\begin{tabular}{c|c|c|}
\cline{2-3} 
& Parameter & Number of bits \\ \hline
\multicolumn{1}{|c|}{\multirow{4}{*}{SISO}} & Received complex input $(x^I_{r,q},x^Q_{r,q})$ & (10,10) \\ \cline{2-3}
\multicolumn{1}{|c|}{} & Coeff. Fading \& Variance  $(h'^2_q)/(\sigma^2)$ & 8 \\ \cline{2-3}
\multicolumn{1}{|c|}{demapper} & Constellation complex symbol $(s^I_{r,j},s^Q_{r,j})$ & (12,12) \\ \cline{2-3}
\multicolumn{1}{|c|}{} & Euclidean distance $A_q$ & 19 \\ \hline

\multicolumn{1}{|c|}{\multirow{3}{*}{SISO}} & Received 4 LLRs & $4 \times 5$ \\ \cline{2-3}
\multicolumn{1}{|c|}{} & Branch metric $\gamma_k$ & 10 \\ \cline{2-3}
\multicolumn{1}{|c|}{\multirow{1}{*}{decoder}} & State metric $\alpha_k,\beta_k$ & 10 \\ \cline{2-3}
\multicolumn{1}{|c|}{} & Extrinsic information $z$ & 10 \\ \hline
\end{tabular}
}
\caption{Typical quantization values.}
\label{OperBit} %\vspace{-10pt}
}
\hfill
\parbox{.5\linewidth}{
\scalebox{0.85}{
\begin{tabular}{|c|c|}
%\centering
\hline
Arithmetic operations ($n_2 \ge n_1$) & Normalized arithmetic operations \\
\hline
1 $Add(n_1,n_2)$ & $0.5 \times (n_1+n_2-1)$$Add(1,1)$ \\
\hline
1 $Sub(n_1,n_2)$ & $0.5 \times (n_1+n_2)$$Add(1,1)$ \\
\hline
1 $Mul(n_1,n_2)$ & [$(n_1-1)(n_2-1)+1-0.5 \times n_1$]$Add(1,1)$ \\
\hline
\end{tabular}
}
\caption{Complexity normalization in terms of $Add(1,1)$.}% when $n_2 \ge n_1$}.
\label{TableApproxGate}
}
\end{table}

%\begin{table}[h!]
%\centering
%\scalebox{0.9}{
%\begin{tabular}{c|c|c|}
%\cline{2-3} 
%& Parameter & Number of bits \\ \hline
%\multicolumn{1}{|c|}{\multirow{4}{*}{SISO}} & Received complex input $(x^I_{r,q},x^Q_{r,q})$ & (10,10) \\ %\cline{2-3}
%\multicolumn{1}{|c|}{} & Coeff. Fading \& Variance  $(h'^2_q)/(\sigma^2)$ & 8 \\ \cline{2-3}
%\multicolumn{1}{|c|}{demapper} & Constellation complex symbol $(s^I_{r,j},s^Q_{r,j})$ & (12,12) \\ \cline{2-3}
%\multicolumn{1}{|c|}{} & Euclidean distance $A_q$ & 19 \\ \hline

%\multicolumn{1}{|c|}{\multirow{3}{*}{SISO}} & Received 4 LLRs & $4 \times 5$ \\ \cline{2-3}
%\multicolumn{1}{|c|}{} & Branch metric $\gamma_k$ & 10 \\ \cline{2-3}
%\multicolumn{1}{|c|}{\multirow{1}{*}{decoder}} & State metric $\alpha_k,\beta_k$ & 10 \\ \cline{2-3}
%\multicolumn{1}{|c|}{} & Extrinsic information $z$ & 10 \\ \hline
%\end{tabular}
%}
%\caption{Typical quantization values.}
%\label{OperBit} \vspace{-20pt}
%\end{table}

\begin{table*}[t!]
\centering
\scalebox{0.75}{
\begin{tabular}{|p{1.7cm}|c|c|}
\hline

\multirow{4}{1.7cm}{SISO rotated demapper with {\it a priori} input} &  Computation units & Number and Type of operations per modulated symbol per turbo demapping iteration \\ \cline{2-3}
	& Euclidean distance 		& $2^M$$Add(18,18)$ + $2^{M+1}$$Sub(8,10)$ + $2^{M+1}$$Mul(8,8)$ + $2^{M+1}$$Mul(8,10)$ + 2$load(10)$ + $(1+2^M)$$load(8)$ \\ \cline{2-3}
	& \multirow{2}{*}{{\it a priori} adder}  			& $(2^M-2)$\{$E[\frac{M-1}{2}]$$Add(8,8)$ + $E[\frac{M-1}{4}]$$Add(9,9)$ + $E[\frac{M-1}{8}]$$Add(10,10)$ + $M$$Sub(8,11)$ + $M$$Sub(11,19)$\} + $M$$load(8)$ + $2^M$$load(M)$ \\ 
	& & For QPSK $M*2^M-2$$Sub(11,19)$ + $M$$load(8)$ + $2^M$$load(M)$ \\ \cline{2-3}
	& Minimum finder			 	& $M$$Sub(8,8)$ + $M.2^M$$Sub(8,19)$ + $M$$store(8)$ \\ \hline \hline

\multirow{4}{1.7cm}{SISO double binary turbo decoder} &  Computation units & Number and Type of operations per coded symbol per turbo decoding iteration \\ \cline{2-3}
	& Branch metric 				& 4$Add(5,5)$ + 38$Add(5,10)$ + 4$Sub(5,5)$ + 8$load(5)$ + 6$load(10)$ \\ \cline{2-3}
	& State metric  				& 64$Add(10,10)$ + 48$Sub(9,9)$ + 8$store(10)$ \\ \cline{2-3}
	& extrinsic information & 32$Add(10,10)$ + 32$Sub(9,9)$ + 9$Sub(10,10)$ + 3$Mul(4,10)$ + 8$load(10)$ + 5$store(10)$ \\ \hline 

\end{tabular}
}
\caption{Complexity computation summary.}
\label{ComplexitySummary} \vspace{-30pt}
\end{table*}

\vspace{-5pt}
\subsection{Complexity evaluation of SISO demapper}
%The complexity of SISO demapping depends on the modulation order (in the context of the above fixed parameters). In fact, for each received modulated symbol $x_{r,q}$ composed of $M$ coded bits, $2^M$ Euclidean distances are computed. With iterative demapping, {\it a priori} information coming from the decoder should be added to the associated Euclidean distance. The minimum distance finder is then applied to search for the closest symbol between the $2^M$ constellation symbols. Thus, the SISO demapper complexity is composed of 3 principal units: Euclidean distance, {\it a priori} adder, and minimum finder functions. For each of these functions we will now consider the equations of subsection \ref{DemapAlg} to compute: (1) the required number and type of arithmetic computations and (2) the required number of read memory access ($load$) and write memory access ($store$). The result of this evaluation is summarized in Table \ref{ComplexitySummary} and explained below. We use the following notation $operation$({\footnotesize NbOfBitsOfOperand1,NbOfBitsOfOperand2}) for arithmetic operations, and $load$({\footnotesize NbOfBits})/$store$({\footnotesize NbOfBits}) for read/write memory operations. Thus, $add(8,10)$ indicates an addition operation of two operands; one quantized on 8 bits and the second on 10 bits. Similarly, $load(8)$ indicates a read access memory of 8-bit word length. 
The complexity of SISO demapping depends on the modulation order (in the context of the above fixed parameters). We will now consider the equations of subsection \ref{DemapAlg} to compute: (1) the required number and type of arithmetic computations and (2) the required number of read memory access ($load$) and write memory access ($store$). The result of this evaluation is summarized in Table \ref{ComplexitySummary} and explained below. We use the following notation $operation$({\footnotesize NbOfBitsOfOperand1,NbOfBitsOfOperand2}) for arithmetic operations, and $load$({\footnotesize NbOfBits})/$store$({\footnotesize NbOfBits}) for read/write memory operations. Thus, $add(8,10)$ indicates an addition operation of two operands; one quantized on 8 bits and the second on 10 bits. Similarly, $load(8)$ indicates a read access memory of 8-bit word length. 

\begin{enumerate}
\item \textbf{Euclidean distance computation} \\
For each modulated symbol (input of the demapper):
\begin{itemize}
\item One $load(8)$ to access the fading channel coefficient normalized by the channel variance $\frac{h'^2_q}{\sigma^2}$ 
\item Two $load(10)$ to access the channel symbols $x^I_{r,q}$ and $x^Q_{r,q}$.
\item For each one of the $2^M$ symbols of the constellation ($s^I_{r,j}$ , $s^Q_{r,j}$):
\begin{itemize}
\item Two $load(8)$ to access the constellation symbols $s^I_{r,j}$ and $s^Q_{r,j}$
\item Two $Sub(8,10)$ to compute ($x^I_{r,q}-s^I_{r,j}$) and ($x^Q_{r,q}-s^Q_{r,j}$)
\item Two $Mul(8,10)$ to multiply with the channel coefficients $\frac{h'_q}{\sigma}$ and $\frac{h'_{q-1}}{\sigma}$
\item Two $Mul(10,10)$ to compute the square of the results above
\item One $Add(18,18)$ to realize the sum of the two Euclidean distance terms
\end{itemize}
\end{itemize}

\item \textbf{ {\it \textbf{A priori}} adder} \\
For each modulated symbol (input of the demapper):
\begin{itemize}
\item $M$ $load(8)$ to access the {\it a priori} informations $L_{apr,Dem}(c_{p,q})$
\item For each one of the $2^M$ symbols of the constellation ($s^I_{r,j}$ , $s^Q_{r,j}$), except two symbols corresponding to all zeros and all ones:
\begin{itemize}
\item One $load(M)$ to access constellation symbol bits $c_{p,q}. \ k=0, 1, \ldots, M-1$
\item $E[\frac{M-1}{2}]$ $Add(8,8)$ to realize the sum of two $L_{apr,Dem}(c_{p,q})$
\item $E[\frac{M-1}{4}]$ $Add(9,9)$ to realize the sum of four $L_{apr,Dem}(c_{p,q})$
\item $E[\frac{M-1}{8}]$ $Add(10,10)$ to realize the sum of eight $L_{apr,Dem}(c_{p,q})$
\item $M$ $Sub(8,11)$ to subtract the LLR of the specific $p^{th}$ bit and thus obtain $B_{p,q}$
\item $M$ $Sub(11,19)$ to realize $A_q-B_{p,q}$ 
\end{itemize}
$E[x]$ represents here the ordinary rounding  of the positive number $x$ to the nearest integer.
\end{itemize}

However, for the simple QPSK modulation the above operations can be simplified as only 2 LLRs exist for one modulated symbol. In fact, in equation (\ref{APrioriAdder}) there is no need to execute an addition followed by a subtraction of the same LLR. Thus, the total number of required arithmetic operations in this case is $4Sub(11,19)$.   

\item \textbf{Minimum finder} \\
For each one of the $M$ bits per modulated symbol:
\begin{itemize}
\item $2^M$ $Sub(19,19)$ to realize the two min operations of equation (\ref{DemapperEquationMaxLogMAP})
\item One $Sub(8,8)$ to subtract the above found 2 minimum values resulting in the demapper extrinsic information
\item One $store(8)$ to store the extrinsic information value
\end{itemize}
\end{enumerate}

\vspace{-5pt}
\subsection{Complexity evaluation of SISO decoder}
The SISO decoder complexity is composed of 3 principal units: branch metric, state metric, and extrinsic information functions. As for the SISO demapper, the result of the complexity evaluation is summarized in Table \ref{ComplexitySummary} and explained below. As stated before, the considered turbo code is an 8-state double binary one. At the turbo decoder side, each double binary symbol should be decoded to take a decision over the 4 possible values $(00, 01, 10, 11)$.
\begin{enumerate}
\item \textbf{Branch metrics ($\gamma$)} \\
For each coded symbol (input of the decoder):
\begin{itemize}
\item 4 $load(5)$ to access systematic and parity LLRs
\item 3 $load(10)$ to access demapper normalized extrinsic informations 
%$Ext_{01}$, $Ext_{10}$ and  $Ext_{11}$
\item 2 $Add(5,5)$ and 2 $Sub(5,5)$ to compute systematic and parity branch metrics $\gamma^{Sys}_{11}$, $\gamma^{Sys}_{10}$, $\gamma^{Parity}_{11}$ and  $\gamma^{Parity}_{10}$
\item 19 $Add(5,10)$ to compute branch metrics $\gamma_k$ and $\gamma^{Sys}_{k}+\gamma^{Parity}_{k}$
\end{itemize}
Operations above should be multiplied by 2 to generate forward and backward branch metrics.

\item \textbf{State metrics ($\alpha$,$\beta$)} \\
For each coded symbol (input of the decoder):
\begin{itemize}
\item 32 $Add(10,10)$ to compute $\alpha_{k-1}(s^{'}) + \gamma_{k}(s^{'},s)$ for the 32 trellis transitions (8-state double binary trellis)
\item 24 $Sub(9,9)$ to realize the 8 max (4-input) operations of equation (\ref{AlfaComput}). In fact, 1 max (N-input) can be implemented as N-1 max (2-input) operations. 1 max (2-input) corresponds to 1 $Sub$
\item 8 $store(10)$ to store computed state metrics only for left butterfly algorithm
\end{itemize}
Operations above should be multiplied by 2 to generate forward $\alpha$ and backward $\beta$ state metrics.

\item \textbf{Extrinsic information ($z$)} \\
For each coded symbol (input of the decoder): 

\begin{itemize}
\item 8 $load(10)$ to access state metric values
\item 32 $Add(10,10)$ to compute the second required addition operation in equation (\ref{SoftOutput}) for the 32 trellis transitions
\item 28 $Sub(9,9)$ to realize the 4 max (8-input) operations of equation (\ref{SoftOutput})
\item 4 $Sub(10,10)$ to subtract symbol-level intrinsic information from the computed soft value (generating symbol-level extrinsic information)
\item 8 $Sub(9,9)$ and 4 $Sub(10,10)$ to realize the 8 max (2-input) operations and compute 4 bit-level (systematic and parity) extrinsic information as demapper {\it a priori} information (equations (\ref{ExtrinsicBitOutput1}) and (\ref{ExtrinsicBitOutput2})). This computation is done only for one of the two SISO decoders
\item 4 $store(10)$ to store the computed bit-level (systematic and parity) extrinsic information
\item 3 $Sub(10,10)$ to normalize symbol-level extrinsic information by subtracting the one related to decision $00$
\item 3 $Mul(4,10)$ to multiply the symbol-level extrinsic information by a scaling factor SF
\item 3 $store(10)$ to store the computed $\mathtt{DEC_1}$ symbol-level extrinsic information as $\mathtt{DEC_2}$ {\it a priori} information
\end{itemize}
\end{enumerate}

\vspace{-5pt}
\subsection{Complexity normalization}
The above conducted complexity analysis exhibits different arithmetic and memory operation types and operand sizes. In order to provide a fair evaluation of the improvement in complexity and memory access with the technique proposed in section \ref{RedDemIterSection}, complexity normalization is necessary. 

For arithmetic operations, normalization can be done in terms of 2-input one bit full adders ($Add(1,1)$). Each one of the adders, subtractors, and multipliers can be converted into an equivalent number of $Add(1,1)$. For adders and subtractors, bit-to-bit half and full adders are used and generalized for operand sizes $n_1$ and $n_2$. Obtained formulas are summarized in Table \ref{TableApproxGate} with simple, yet accurate, analysis of all corner cases. Similarly, multiplication operations are normalized using successive addition operations. Memory access operation of $m$ word of size $n$ are normalized to one memory access operation of $m\times n$ bits.

%\begin{table}[h!]
%\centering
%\scalebox{0.85}{
%\begin{tabular}{|c|c|}
%\centering
%\hline
%Arithmetic operations ($n_2 \ge n_1$) & Normalized arithmetic operations \\
%\hline
%1 $Add(n_1,n_2)$ & $0.5 \times (n_1+n_2-1)$$Add(1,1)$ \\
%\hline
%1 $Sub(n_1,n_2)$ & $0.5 \times (n_1+n_2)$$Add(1,1)$ \\
%\hline
%1 $Mul(n_1,n_2)$ & [$(n_1-1)(n_2-1)+1-0.5 \times n_1$]$Add(1,1)$ \\
%\hline
%\end{tabular}
%}
%\caption{Complexity normalization in terms of $Add(1,1)$.}% when $n_2 \ge n_1$}.
%\label{TableApproxGate}
%\end{table}
%\vspace{-20pt}
Applying the proposed complexity normalization approach to Table \ref{ComplexitySummary} leads to the results summarized in Table \ref{ComplexitySummaryNormlization}.

\vspace{-5pt}
\section{Discussions and achieved gains}
\label{DiscussionSection}
This section evaluates and discusses the achieved complexity reductions using the proposed original iteration scheduling of TBICM-ID-SSD at different modulation orders and code rates. As concluded in section \ref{RedDemIterSection}, two demapping iterations can be eliminated while keeping the number of turbo decoding iterations unaltered. Overall, this will lead to a reduction corresponding to two times the execution of the SISO demapping function. Besides the fact that the obtained results will depend on the modulation order and code rate, a third parameter should be considered regarding the iterative demapping implementation choice. In this regard, two configurations should be analyzed. In the first configuration, denoted CASE 1, the Euclidean distances are re-calculated at each demapping iteration. While in the second configuration, denoted CASE 2, the computation of the Euclidean distances are done only once, at the first iteration, then stored and reused in later demapping iterations. Thus, CASE 1 implies higher arithmetic computations, however less memory access, than CASE 2.

%\vspace{-10pt}
Using the normalized complexity evaluation of Table \ref{ComplexitySummaryNormlization}, achieved gains comparing $4IDem\_2EIDec$ to $6IDem$ for all configurations are summarized in Table \ref{GainComplexityRedIter}. In the following we will explain first how these values are computed and then discuss the obtained results.

In fact, considering the code rate $R_c$ and the number of bits per symbol $M$, the relation between the number of double binary coded symbols ($N_{CodedSymb}$) and the corresponding number of modulated symbols ($N_{ModSymb}$) can be written as follows.

{\small \vspace{-10pt}
\begin{equation}
\label{ConversionExpressionModulCoded}
N_{ModSymb}=\frac{2 . N_{CodedSymb}}{M . R_c}
\end{equation} \vspace{-15pt}
}

The complexity reduction ($G$) corresponds to the ratio between the complexity of two SISO demapping executions and the complexity of the original TBICM-ID-SSD configuration. If the original TBICM-ID-SSD configuration requires $N_{it}$ iterations to process a frame composed of $N_{ModSymb}$ modulated symbols (equivalent to $N_{CodedSymb}$ coded symbols), the complexity reduction can be approximated by the following expression.

{\small %\vspace{-5pt}
\begin{equation}
\label{GainExpression}
G=\frac{2 . F_{Dem}(M) . N_{ModSymb}}{N_{it} . F_{Dem}(M) . N_{ModSymb} + N_{it} . F_{Dec} . N_{CodedSymb}}
\end{equation} %\vspace{-10pt}
}
where $F_{Dem}$ designates the complexity of SISO demapper which depends on the constellation size and $F_{Dec}$ designates the complexity of SISO decoder. When converting in this equation the number of modulated symbols into equivalent coded symbols using equation (\ref{ConversionExpressionModulCoded}), we obtain the following equation.

{\small
\begin{equation}
\label{GainExpressionFinal}
G=\frac{2 . F_{Dem}(M)}{N_{it} . F_{Dem}(M) + N_{it} . F_{Dec} . \frac{M . R_c}{2}}
\end{equation}
}

\begin{table*}[t!]
\centering
\scalebox{0.85}{
\begin{tabular}{|p{1.5cm}|c|c|}
\hline

\multirow{4}{1.7cm}{SISO rotated demapper with {\it a priori} input} &  Computation units & Number and Type of operations per modulated symbol per turbo demapping iteration \\ \cline{2-3}
	& Euclidean distance 		& 123.75.$2^{M+1}$$Add(1,1)$ + $load(28+2^{M+3})$ \\ \cline{2-3}
	& \multirow{2}{*}{{\it a priori} adder}  			& $(2^M-2)$\{7.5$E[\frac{M-1}{2}]$+ 8.5$E[\frac{M-1}{4}]$ + 9.5$E[\frac{M-1}{8}]$ + 24.5.$M$\}$Add(1,1)$ + $load(8M)$ + $load(M2^M)$ \\ 
	& & For QPSK  15.$M(2^M-2)$$Add(1,1)$ + $load(8.M+M.2^M)$ \\ \cline{2-3}
	& Minimum finder			 	& $(8+13.5.2^M)M$$Add(1,1)$ + $store(8.M)$ \\ \hline \hline

\multirow{4}{1.7cm}{SISO double binary turbo decoder} &  Computation units & Number and Type of operations per coded symbol per turbo decoding iteration \\ \cline{2-3}
	& Branch metric 				& 304$Add(1,1)$ + $load(100)$ \\ \cline{2-3}
	& State metric  				& 1040$Add(1,1)$ + $store(80)$ \\ \cline{2-3}
	& extrinsic information & 760$Add(1,1)$ + $load(80)$ + $store(50)$ \\ \hline 

\end{tabular}
}
\caption{Complexity computation summary after normalization.}
\label{ComplexitySummaryNormlization} \vspace{-7pt}
\end{table*}

\begin{table*}[t!]
\centering
\scalebox{0.9}{
\begin{tabular}{|c|c|c|c|c|c|c|c|c|c|c|c|c|}
\hline
Modulation scheme & \multicolumn{6}{c|}{CASE1 (With recomputed Euclidean distances)} & \multicolumn{6}{c|}{CASE2 (With stored Euclidean distances)} \\
\cline{2-13}
 & \multicolumn{3}{c|}{$R_c = 1/2$} & \multicolumn{3}{c|}{$R_c = 6/7$} & \multicolumn{3}{c|}{$R_c = 1/2$} & \multicolumn{3}{c|}{$R_c = 6/7$} \\
\cline{2-13}

 & \multicolumn{3}{c|}{Complexity Reduction} & \multicolumn{3}{c|}{Complexity Reduction} & \multicolumn{3}{c|}{Complexity Reduction} & \multicolumn{3}{c|}{Complexity Reduction} \\
 & $arith$ & $load$ & $store$ & $arith$ & $load$ & $store$ & $arith$ & $load$ & $store$ & $arith$ & $load$ & $store$ \\ \hline

QPSK & $11.9\% $ & $10.6\%$ & $3.7\%$ & $8.2\%$ & $7.1\%$ & $2.2\%$ & $2.5\%$ & $12\%$ & $3.4\%$ & $1.6\%$ & $8.2\%$ & $2.1\%$ \\
\hline
QAM16 & $20.3\%$ & $13.7\%$ & $3.7\%$ & $15.9\%$ & $9.7\%$ & $2.2\%$ & $11.6\%$ & $18.1\%$ & $3.1\%$ & $8.3\%$ & $13.4\%$ & $2\%$ \\
\hline
QAM64 & $27.9\%$ & $21.4\%$ & $3.7\%$ & $25\%$ & $17.1\%$ & $2.2\%$ & $21.8\%$ & $26.5\%$ & $2.5\%$ & $18.5\%$ & $22.3\%$ & $1.7\%$ \\
\hline
QAM256 & $31.6\%$ & $28.4\%$ & $3.7\%$ & $30.5\%$ & $25.7\%$ & $2.2\%$ & $27.6\%$ & $32.2\%$ & $1.5\%$ & $26.2\%$ & $30\%$ & $1.2\%$ \\
\hline
\end{tabular}
}
\caption{Reduction in number of operations, read/write access memory comparing "$4IDem\_2EIDec$" to "$6IDem$" for different modulation schemes and code rates.}
\label{GainComplexityRedIter} \vspace{-20pt}
\end{table*}

\vspace{-10pt}
This last equation has been used to obtain individually the complexity reductions in terms of arithmetic, read memory access, and write memory access operations of Table \ref{GainComplexityRedIter}, for $N_{it}=6$. For CASE 1, results show increased benefits in terms of number of arithmetic operations (up to $31.6\%$) and read memory accesses (up to $28.6\%$) with higher modulation orders. This can be easily predicted from equation (\ref{GainExpressionFinal}) as the value of $F_{Dem}$ increases with the constellation size. The equation shows also that the higher the code rate is, lower the benefits are.

On the other hand, the improvement in write memory access ($3.7\%$ for $R_c=1/2$ and $2.2\%$ for $R_c=6/7$) is low and constant for all modulation orders. In fact, in Table \ref{GainComplexityRedIter} the single memory $store$ term which depends on the modulation order is $store(8.M)$ for the minimum finder computation. This term is required per modulated symbol and when converted to the equivalent number per coded symbol (equation (\ref{ConversionExpressionModulCoded})) for a fixed code rate a constant value, independent from $M$, is obtained.  

Similar behavior is shown for CASE 2, except for two points. The first one concerns the improvements in arithmetic operations and read memory accesses. In fact, compared to CASE 1, this configuration implies less arithmetic and more memory access operations which lead to less benefits for the former and more benefits for the latter (equation (\ref{GainExpressionFinal})). The second point concerns the improvement in write memory access. In fact, besides the term $M \times 8 bits$, a value of $19 \times 2^M$ is required only for the first iteration to store the $2^M$ Euclidean distances quantized on $19$ bits each. This added value is much higher in comparison to the reduced $M \times 8 bits$ write memory access. Therefore the improvement in write access memory operations will be less for higher constellation sizes (down to $1.2\%$). 

\textcolor{black}{It is worth noting that applying the proposed scheduling combined with an early stopping criteria might diminish the benefit from the scheduling, but at the cost of an additional complexity.} 

\section{Conclusion}
\label{Conclusion}
Convergence speed analysis is crucial in TBICM-ID-SSD systems in order to tune the number of iterations to be optimal when considering the practical implementation perspectives. Conducted analysis has demonstrated that omitting two turbo demodulation iterations without decreasing the total number of turbo decoding iterations leads to promising complexity reductions while keeping error rate performance almost unaltered. A maximum loss of 0.15 dB is shown for all modulation schemes and code rates in a fast-fading channel with and without erasure. The number of normalized arithmetic operations is reduced from $8.2\%$ for QPSK configuration to $\frac{2}{N_{it}}\%$ for QAM256 (e.g. for $N_{it}=6$ this gives a reduction of $33.3\%$). Similarly, the number of read access memory is reduced in a range between $8.2\%$ to $\frac{2}{N_{it}}\%$. This complexity reduction improves significantly latency and power consumption, and thus paves the way towards the adoption of TBICM-ID-SSD hardware implementations in future wireless receivers. Future work targets the extension of this analysis to other baseband iterative applications and its integration into available hardware prototypes. 

%\appendices

%\section*{Acknowledgment}

%\begin{thebibliography}{1}
 %0.5em minus 0.4em\relax Harlow, England: Addison-Wesley, 1999.
%\end{thebibliography}

\bibliographystyle{IEEEtran}
\bibliography{MyLibrary}

%\begin{IEEEbiography}{Salim Haddad}
%Salim Haddad received the engineer degree
%in electrical-electronics engineering, option telecommunication from the Universit\'e Libanaise, Facult\'e de G\'enie, Tripoli, Liban, in 2009, and the MS degree in Signaux et Circuits from the Universit\'e de Bretagne Occidentale (UBO), Brest, France, in 2009. He is with ........................His current research interests include iterative receivers, performance estimation and complexity reduction. ......
%\end{IEEEbiography}

%\begin{IEEEbiographynophoto}{Amer Baghdadi}

%\begin{IEEEbiographynophoto}{Michel Jezquel}

%\begin{IEEEbiography}{Salim Haddad}
%\end{IEEEbiography}

%\begin{IEEEbiographynophoto}{Amer Baghdadi}

%\newpage

%\begin{IEEEbiographynophoto}{Michel Jezequel}
\end{document}